\def\F{F}
\def\J{J} 
\def\L{{\mathcal L}}
\def\R{{\mathcal R}}
\def\T{{\mathcal T}}
\def\A{{\mathcal A}}
\def\B{{\mathcal B}}  
\def\C{{\mathcal C}}
\def\D{{\mathcal D}}
\def\X{{\mathcal X}}
\def\i{{\mathrm i}}
\def\d{{\mathrm d}}
\def\e{{\mathrm e}}
\def\dps{\displaystyle}
\def\lsub#1#2{\kern+.2ex {}_#1 \kern-.2ex #2}
\def\vecx{\left( \begin{array}{@{}c@{}} A \\ B \\ C \\ D \end{array} \right)}
\def\vecxt{\left( \begin{array}{@{}c@{}} \hat{A} \\ \hat{B} \\ \hat{C} \\
\hat{D} \\ \hat{X} \end{array}\right)}
\def\l{\langle W|}
\def\lt{\lsub{{\tau_1}}{\langle W|}}
\def\ln{\lsub{0}{\langle W|}}
\def\lo{\lsub{1}{\langle W|}}
\def\sln{\langle {\mathcal W}_0|}
\def\slo{\langle {\mathcal W}_1|}

\def\rt{|V\rangle_{\tau_L}}
\def\rn{|V\rangle_0}
\def\ro{|V\rangle_1}
\def\srn{|{\mathcal V}_0\rangle}
\def\sro{|{\mathcal V}_1\rangle}
\def\zn#1{|#1\rangle\kern-.15em\rangle_0}
\def\zo#1{|#1\rangle\kern-.15em\rangle_1}
\def\ev#1{|#1;#1^{-1}\rangle\kern-.15em\rangle}
\def\eoo{\left(\begin{array}{@{}cc@{}} 1 & 0 \\ 0 & 0
\end{array}\right)}
\def\eot{\left(\begin{array}{@{}cc@{}} 0 & 1 \\ 0 & 0
\end{array}\right)}
\def\eto{\left(\begin{array}{@{}cc@{}} 0 & 0 \\ 1 & 0
\end{array}\right)}
\def\ett{\left(\begin{array}{@{}cc@{}} 0 & 0 \\ 0 & 1
\end{array}\right)}

\def\binom#1#2{\left( \begin{array}{@{}c@{}} #1 \\ #2 \end{array} \right)}

\DeclareFontShape{OML}{cmm}{m}{b}{%
   <-> cmmib10}{}
\DeclareMathAlphabet{\mathbf}{OML}{cmm}{m}{b}
\DeclareSymbolFont{boldletters}{OML}{cmm}{m}{b}
\DeclareMathSymbol{\boldtau}{\mathord}{boldletters}{28}

\documentstyle[pre,aps,array,epsfig]{revtex}
\begin{document}
\draft
\title{Exact Stationary State for an ASEP with Fully Parallel
Dynamics}
\author{Jan de Gier$^{\mathrm a}$ \and Bernard Nienhuis$^{\mathrm b}$}
\address{$^{\mathrm a}$ Institute for Theoretical Physics,
University of Utrecht, Princetonplein 5, 3584 CC Utrecht, The
Netherlands}
\address{ $^{\mathrm b}$  Institute for
Theoretical Physics, University of Amsterdam, Valckenierstraat 65,
1018 XE Amsterdam, The Netherlands}
\date{\today}

\maketitle

\begin{abstract}
The exact stationary state of an asymmetric exclusion process with
fully parallel dynamics is obtained using the matrix product
Ansatz. We give a simple derivation for the deterministic case by a
physical interpretation of the dimension of the matrices.
We prove the stationarity via a cancellation mechanism and by
making use of an explicit representation of the matrix algebra we
easily find closed expressions for the correlation functions in the
general probabalistic case. Asymptotic expressions, obtained by
making use of earlier results, allow us to derive the exact phase
diagram. 
\end{abstract}

\section{Introduction}

In this paper we describe the exact stationary state
of a asymmetric exclusion process (ASEP) with fully parallel
dynamics and open boundaries. This is a special case of the
Nagel-Schreckenberg model for traffic flow
\cite{Nagel:1992,Schreck:1995}. Exact results have been known for
some time for several update rules such as random  sequential
\cite{Derrida:1992,Schuetz:1993} and sublattice parallel
\cite{Kandel:1990,Schuetz:1993b,Hinrichsen:1996,Honecker:1997}. For
fully parallel dynamics and open boundary conditions mean field
results have been obtained recently
\cite{Rajewsky:1998,Tilstra:1998}. In a recent preprint,
Evans {\em et al.} \cite{Evans:1998} presented an exact solution of
this model using a site oriented matrix product Ansatz
\cite{Derrida:1993}. Using an exlicit representation of the resulting
algebra they calculated the current and density profile via 
generating function techniques. 

We will present a simple and physical derivation of the solution for
deterministic bulk dynamics. This solution leads us to a bond oriented
matrix product Ansatz resulting in a matrix algebra for stochastic bulk
dynamics. Using an explicit representation of this algebra it is shown
that difficult recursion relations can be circumvented and that
integral expressions for the current and density profile can be given
almost immediately. The resulting integrals can be calculated
resulting in closed expressions similar to those of the random
sequential case.  

The outline of the paper is as follows. In section \ref{se:model} the
model is defined and some notation is fixed. In section \ref{se:recur}
we find and solve a simple recursion relation for the deterministic
case. This solution can be naturally recast in the form of a matrix
product through an interpretation of the dimension of the matrices,
which is done in section \ref{se:matprod}. Section \ref{se:algebra} is
concerned with the formulation of a matrix product Ansatz for the
general case of which solutions are presented in section
\ref{se:rep}. In the subsequent section it is shown that with a 
diagonalization procedure closed expressions for the density profile and
other correlation functions are obtained easily. Finally, in section
\ref{se:phase}, the phase diagram is derived using asymptotic
expression for the density profile and the current.

\section{The model}
\label{se:model}

The model is defined on a one dimensional lattice with $L$ sites. Each
site may be occupied with a particle or it may be
empty. Configurations on the lattice are written as
$\{\tau\}=\{\tau_1,\ldots,\tau_L\}$ numbering from left to right, 
where the presence or absence of a
particle at site $i$ is denoted by $\tau_i=1$ or $\tau_i=0$. An
asymmetric exclusion process (ASEP) is defined by imposing the
following dynamics at each timestep $t \rightarrow t+1$ for the
particles: If there is a particle at site $i$ and if site $i+1$ is
empty, it hops to site $i+1$ with probability $p$ and remains at site
$i$ with probability $1-p$. If site $i+1$ is occupied the particle at
site $i$ remains there with probability $1$. This dynamics is applied
to all particles at the same time, hence the name fully
parallel. This dynamics is also known as the rule-184 cellular
automaton which prescribes how the value of $\tau_i$ at time $t+1$
depends on the values of $\tau_{i-1}$, $\tau_i$ and $\tau_{i+1}$ at
time $t$. Given the configuration $\{\tau\}$ at time $t$, the
configuration $\{\tau'\}$ at time $t+1$ is given by 
\begin{equation}
\tau'_i = \hat{p}_{i-1}\tau_{i-1}(1-\tau_i) + (1-\hat{p}_i)
\tau_i(1-\tau_{i+1}) + \tau_i\tau_{i+1},
\quad(i=2,\ldots,L-1), \label{eq:bulkdyn}   
\end{equation} 
where \(\hat{p}_i\) are stochastic boolean variables with mean \( \langle 
\hat{p}_i\rangle = p\).
The lattice is coupled to two reservoirs at the first and last
sites. Particles may enter the system from the first reservoir if the
first site is empty with rate $\alpha$ and they may leave the system
with rate $\beta$ into the second reservoir at the last site. Thus,
\begin{eqnarray}
\tau'_1 &=& \hat{\alpha} (1-\tau_1) + (1-\hat{p}_1) \tau_1 (1-\tau_2) +
\tau_1 \tau_2, \label{eq:bound1}\\ 
\tau'_L &=& \hat{p}_{L-1}\tau_{L-1} (1-\tau_L) + (1-\hat{\beta})
\tau_L, \label{eq:bound2}
\end{eqnarray} 
where $\hat{\alpha}$ and $\hat{\beta}$ are boolean
variables such that $\langle \hat{\alpha} \rangle = \alpha$ and $\langle
\hat{\beta} \rangle = \beta$. Note
that the dynamical rules have a manifest particle hole symmetry given
by
\begin{equation}
\begin{array}{@{}>{\dps}rc>{\dps}l@{}}
\tau_i &\rightarrow & 1-\tau_{L-i+1},\\
\hat{\alpha} &\leftrightarrow & \hat{\beta}.
\end{array} \label{eq:phsym}
\end{equation}

Let us now associate to every configuration $\{\tau\}$ a vector
$\boldtau$ defining an orthonormal basis of a Hilbert
space. A state ${\mathbf P}(t)$ of the system will be any vector in this
Hilbert space, i.e.
\begin{equation}
{\mathbf P}(t) = \sum_{\{\tau\}} P_t(\tau_1,\ldots,\tau_L) \boldtau.
\label{eq:state}
\end{equation}
The time evolution of such a state may be written as follows,
\begin{equation}
{\mathbf P}(t+1) = {\mathbf T} {\mathbf P}(t), \label{eq:evol}
\end{equation}
where ${\mathbf T}$ is called the transfer matrix. First of all one
would like to know the stationary state of this model, i.e. the
eigenstate of the transfer matrix with eigenvalue 1. As this state is
time independent we will suppress from here on 
the temporal suffix of \(P\).

\section{Recurrence relation}
\label{se:recur}

For the deterministic bulk dynamics $(p=1)$ the bulk relations
(\ref{eq:bulkdyn}) simplify. Tilstra and Ernst \cite{Tilstra:1998}
have shown that in this case each configuration that can occur
in the stationary state may be spatially divided into three parts, a free
flow part, a jammed flow part and an interface of varying width.
The free flow part is
defined to be that part of the configuration up to the rightmost $00$ pair
and consists of isolated particles only. The jammed flow part starts
with the leftmost $11$ pair and  
contains isolated holes only. The jammed flow and the free flow can not 
overlap, but they may be separated by an interface 
consisting of a sequence of $10$ pairs. Using this
identification the dynamical rules become very simple. Denote the site
of the last zero of the last $00$ pair by $f$ and the site of the
first one of the first $11$ pair by $j$, then the rules are given by
\begin{equation}
\begin{array}{@{}>{\dps}rc>{\dps}l>{\dps}rc>{\dps}lr@{}}
\tau'_i &=& \tau_{i-1}, &&& \quad (i=2,\ldots,f),\\
\tau'_i &=& \tau_{i+1}, &&& \quad (i=j,\ldots,L-1),\\
\tau'_i &=& \tau_{i-1} & = & \tau_{i+1}, & \quad (i=f+1,\ldots,j-1). 
\end{array} \label{eq:bulksimp}
\end{equation}
If there are no $00$ pairs in a particular configuration we set $f=1$
if $\tau_1=0$, otherwise $f=0$. Similarly, if there are no $11$ pairs,
$j=L$ if $\tau_L=1$, otherwise $j=L+1$. In these cases the
bulk relations (\ref{eq:bulksimp}) are not valid but one has to use
the boundary relations (\ref{eq:bound1}) and (\ref{eq:bound2}). The
equations of motion (\ref{eq:evol}) for deterministic bulk
dynamics $(p=1)$ induces for the stationary state the equation
\begin{eqnarray}
\lefteqn{
P(\tau_1,\ldots,\tau_f,(10)^n,\tau_j,\ldots\tau_L) =
\F(\tau_1|\tau_2) \J(\tau_L|\tau_{L-1}) \; \times }
\nonumber \\
&&\quad \left[ 
P(\tau_2,\ldots,\tau_f,(10)^{n+1},\tau_j,\ldots\tau_{L-1})
 + \sum_q
P(\tau_2,\ldots,\tau_f,(10)^q01(10)^{n-q},\tau_j,\ldots\tau_{L-1})
\right].
\label{eq:expl}
\end{eqnarray}
Here, $\F$ and $\J$ are the transition rates for particles
entering and leaving the system. They are given by
\begin{equation}
\begin{array}{@{}>{\dps}rc>{\dps}l>{\dps}rc>{\dps}l@{}}
\F(0|0) &=& 1-\alpha,&\quad \J(1|1) &=& 1-\beta,\\
\F(1|0) &=& \alpha,&\quad \J(0|1) &=& \beta,\\
\F(0|1) &=& 1, &\quad \J(1|0) &=& 1,\\
\F(1|1) &=& 0, &\quad \J(0|0) &=& 0.
\end{array}
\end{equation}
Iteration of equation (\ref{eq:expl}) suggests the following Ansatz for the
probabilities $P$, 
\begin{eqnarray}
P(\tau_1,\ldots,\tau_f,(10)^n,\tau_j,\ldots,\tau_L) &=& \frac{1}{Z_L}
P_{\mathrm f} (\tau_1,\ldots,\tau_f) P_{\mathrm I}(n) P_{\mathrm j} (\tau_j,
\ldots, \tau_L), \label{eq:PAnsatz} \\
P_{\mathrm f} (\tau_1,\ldots,\tau_f) &=& x^f \prod_{i=1}^{f-1}
\F(\tau_i |\tau_{i+1}), \label{eq:Pfdef} \\
P_{\mathrm j} (\tau_j,\ldots,\tau_L) &=& y^{L-j+1} \prod_{i=j}^{L-1}
\J(\tau_{i+1}|\tau_i), \label{eq:Pjdef} 
\end{eqnarray}
where $x$, $y$ and $P_{\mathrm I}(n)$ are to be determined and $Z_L$ is a
normalization. Substituting this Ansatz into (\ref{eq:expl}) we find
that $P_{\mathrm I}(n)$ obeys the following recursion relation, 
\begin{equation}
P_{\mathrm I}(n) = x^{-1} y^{-1} P_{\mathrm I}(n+1) +
(1-\alpha)(1-\beta) \sum_{p=0}^n (\alpha x^2)^p (\beta y^{2})^{n-p},
\label{eq:recur} 
\end{equation}
where we have set $P_{\mathrm I}(0)=1$. Explicit consideration of the
equations of motion for $P(0\ldots 0)$ and $P(0\ldots 01)$ determines
the ratio between $x$ and $y$. A convenient choice that fulfills
this relation is $x=\beta$, $y=\alpha$. The recursion relation
(\ref{eq:recur}) can be solved easily and all the sums can be
performed to give 
\begin{equation}
P_{\mathrm I}(n) = (\alpha \beta)^n \frac{(1-\alpha)\beta^{n+1} -
(1-\beta)\alpha^{n+1}}{\beta-\alpha}. \label{eq:Psol}
\end{equation}
We thus have calculated the complete probability distribution function
for the stationary state for deterministic bulk dynamics. The
normalization in (\ref{eq:PAnsatz}) is given by 
\begin{equation}
Z_L = \frac{(1-\alpha^2) \beta^{L+1} - (1-\beta^2)
\alpha^{L+1}}{\beta-\alpha}.
\end{equation}
All expressions remain valid for
$\alpha=\beta$ by taking the appropriate limits.

\section{Matrix Product}
\label{se:matprod}

We now will construct a matrix product representation of the
stationary state. This will turn out to be useful for calculational
reasons but it also helps us to find the solution for $p <
1$. First the vectors $\boldtau$ are written as product states, 
\begin{equation}
\boldtau = \bigotimes_{i=1}^{L-1}
{\mathbf v}(\tau_i,\tau_{i+1}). \label{eq:tauprod}
\end{equation}
Next we will show how the form of the solution obtained in the
previous section hints in the right direction.

The exact form of $P_{\mathrm I}(n)$ as given by equation (\ref{eq:Psol})
suggests another view at the structure of the probability distribution
function. This becomes clear by rewriting (\ref{eq:Psol}) as
\begin{equation}
P_{\mathrm I}(n) = \sum_{k=0}^n P_0 (2k) - (\alpha \beta)^{1/2} 
\sum_{k=1}^n P_0(2k-1),
\label{eq:PIsum}
\end{equation}
where $P_0$ is given by
\begin{eqnarray}
P_0 (k) &=& (xy)^{-1} P_{\mathrm f} (\tau_f,\ldots,\tau_{f+k})
P_{\mathrm j} (\tau_{f+k+1},\ldots,\tau_{j}), \nonumber\\
&=& (\alpha \beta)^n \alpha^{n-k/2} \beta^{k/2}.
\end{eqnarray}
$P_{\mathrm f}$ and $P_{\mathrm j}$ are defined in equations
(\ref{eq:Pfdef}) and (\ref{eq:Pjdef}). Thus equation (\ref{eq:PIsum})
tells us that the weight of the interface of width $2n$ is equal to a
sum over the positions of a separator. Configurations to the left of
this separator are regarded as belonging to the free flow part and
configurations to the right as belonging to the jammed part. There are
different prefactors for even and odd positions of the separator. 
After equation (\ref{eq:PIsum}) is substituted into equation 
(\ref{eq:PAnsatz}) one may even place the separator inside the free 
flow or the jammed flow, as the resulting additional terms vanish.
This suggests that the stationary state may be written as a matrix 
product,
\begin{equation}
{\mathbf P} = \sum_{\{\tau\}} P(\tau_1,\ldots,\tau_L) \boldtau =
\l \left( \begin{array}{@{}cc@{}} {\mathbf F} & {\mathbf S} \\
{\mathbf 0} & {\mathbf J} \end{array}\right)^{L-1} |V\rangle
/Z_L, \label{eq:matprod} 
\end{equation}
which is to be understood as a normal matrix multiplication but where
the matrix elements are tensored. The matrices ${\mathbf F}$ and 
${\mathbf J}$ govern the
free and jammed flow respectively, while ${\mathbf S}$ is a matrix for the
separator. Indeed, we find
\begin{eqnarray}
{\mathbf F} &=& \left (\begin{array}{@{}cc@{}}
x \F(0|0) {\mathbf v}(00) & x \F(0|1) {\mathbf v}(01) \\
x \F(1|0) {\mathbf v}(10) & 0
\end{array}\right) \;\;=\;\;
\left (\begin{array}{@{}cc@{}}
\beta(1-\alpha) {\mathbf v}(00) & \beta {\mathbf v}(01) \\
\alpha\beta {\mathbf v}(10) & 0
\end{array}\right),\\
{\mathbf J} &=& \left (\begin{array}{@{}cc@{}}
0 & y \J(1|0) {\mathbf v}(01) \\
y \J(0|1) {\mathbf v}(10) & y \J(1|1) {\mathbf v}(11) \\
\end{array}\right) \;\;=\;\;
\left (\begin{array}{@{}cc@{}}
0 & \alpha {\mathbf v}(01) \\
\alpha\beta {\mathbf v}(10) & \alpha(1-\beta) {\mathbf v}(11)
\end{array}\right),\\
{\mathbf S} &=& \left (\begin{array}{@{}cc@{}}
0 & \alpha\beta{\mathbf v}(01) \\
-(\alpha\beta)^2 {\mathbf v}(10) & 0
\end{array}\right),\\
\l &=& (1,1,0,\alpha),\quad \langle V| \;\;=\;\;
(\beta,0,1,1). 
\end{eqnarray} 
In other words, the matrix operator in (\ref{eq:matprod}) has two types 
of binary indices; the explicit ones referring to the type of flow, free or
jammed, are contracted, and the internal ones, the occupation 
numbers \(\tau_i\), are tensored.

\section{Algebra}
\label{se:algebra}

Having found the stationary state for {\em deterministic} 
bulk dynamics (\(p=1\)) 
and its representation as a matrix product, we now reverse the problem and
show that the stationary state may be found by using the matrix
product Ansatz (MPA) technique also for {\em stochastic} (\(p\neq 1\))
dynamics. In the following we will derive a
matrix algebra from the MPA for the ASEP, i.e. for arbitrary $p$. 
We will show that a finite dimensional representation for this algebra
can be found for $p=1$ by using the solution found in section
\ref{se:matprod}. 

As usual for the MPA, the matrix product (\ref{eq:matprod}) will be
written as follows,
\begin{equation}
{\mathbf P} =  \lt \vecx^{\otimes (L-1)} \rt, \label{eq:mpa}
\end{equation}
where $(A,B,C,D)$ is a vector on the basis ${\mathbf v}(00)$,
${\mathbf v}(01)$, ${\mathbf v}(10)$,  ${\mathbf v}(11)$ with matrix
valued entries. It is to be understood that each entry of the tensor
product is bra-ketted with $\lt$ and $\rt$. 
In equation (\ref{eq:mpa}) the tensored indices are explicit, while 
the contracted indices are implicit, this in contrast to (\ref{eq:matprod}).
In the following we will
show that the transfer matrix for the ASEP with fully parallel update
rules may be written as ${\mathbf T} = \R \T^{L-1} \L$, for which the
following mechanism ensures that (\ref{eq:mpa}) is a stationary state,
\begin{eqnarray}
\lt\L\vecx = \lt\vecxt, && \R\vecxt\rt = \vecx\rt, \label{eq:boundalg}\\
\T\left[ \vecxt\otimes\vecx \right] &=&
\vecx\otimes\vecxt. \label{eq:bulkalg}
\end{eqnarray}
In order to find $\L$, $\R$ and $T$ we introduce the following
probability distribution, 
\begin{equation}
P(\tau_1,\ldots,\tau_{i-1};\sigma_{i},\tau_{i+1}\ldots,\tau_L),
\label{eq:partprob} 
\end{equation}
which corresponds to a partial updated sequence with
$\tau_1,\ldots,\tau_{i-1}$ at time $t+1$ and $\tau_{i+1},\ldots,\tau_L$ at
time $t$. 
The variable $\sigma_{i}$ can attain three different values, say $0,1,2$. 
The value $0$ corresponds to a hole both at time $t$ and $t+1$, 
the value $1$ to a particle at time $t$, 
and the value 2 to a particle moving into site $i$ at time $t+1$. 
The introduction of such a third state $\sigma_i=2$ was originally
done by Rajewsky {\em et al} \cite{Rajewsky:1998} and
is necessary to incorporate correctly the fully parallel
update rule. The probabilities (\ref{eq:partprob}) correspond with the
matrix product Ansatz in the following way
\begin{equation}
P(\ldots,\tau_{i-1};\sigma_{i},\tau_{i+1},\ldots) = \lt \cdots
Y(\tau_{i-2},\tau_{i-1}) \hat{Y}(\tau_{i-1},\sigma_{i})
Y(\tau_{i},\tau_{i+1}) \cdots\rt, 
\end{equation}
where $\tau_{i} = \sigma_{i}$ mod 2. This also shows why we need
the fifth matrix $\hat{X}=\hat{Y}(0,2)$ for the intermediate states in
(\ref{eq:boundalg}) and (\ref{eq:bulkalg}). The equations, which
define ${\mathbf T}$, for the probabilities (\ref{eq:partprob}) are explicitly
given by
\begin{equation}\label{eq:Pevol}
\begin{array}{rcl}
P(\ldots,0;0,\ldots)&=&     P(\ldots;0,0,\ldots),\\
P(\ldots,0;1,\ldots)&=&     P(\ldots;0,1,\ldots),\\
P(\ldots,0;2,\ldots)&=&    pP(\ldots;1,0,\ldots),\\
P(\ldots,1;0,\ldots)&=&(1-p)P(\ldots;1,0,\ldots) + P(\ldots;2,0,\ldots),\\
P(\ldots,1;1,\ldots)&=&     P(\ldots;1,1,\ldots) + P(\ldots;2,1,\ldots),\\
P(\ldots,1;2,\ldots)&=& 0
\end{array}
\end{equation}

These equations immediately determine the matrix $\T$.
In a similar fashion the boundary operators $\L$ and $\R$ can be
calculated and they are given by
\begin{equation}\label{eq:L&R}
\L = \left( 
\begin{array}{@{}cccc@{}}
1-\alpha & 0 & 0 & 0 \\
0 & 1-\alpha & 0 & 0 \\
\alpha & 0 & 1-p & 0 \\
0 & \alpha & 0 & 1\\
0 & 0 & p & 0
\end{array}\right),\quad
\R = \left( 
\begin{array}{@{}ccccc@{}}
1 & \beta & 0 & 0 & 0 \\
0 & 1-\beta & 0 & 0 & 1 \\
0 & 0 & 1 & \beta & 0 \\
0 & 0 & 0 & 1-\beta & 0
\end{array}\right)
\end{equation}
With these definitions, equations (\ref{eq:bulkalg}) and (\ref{eq:Pevol})
imply the following algebra,
\begin{equation}
\begin{array}{l}
A\hat{A} = \hat{A}A,\quad  A\hat{B} = \hat{A}B,\quad A\hat{X} =
p\hat{B}C,\quad B\hat{C} = (1-p)\hat{B}C+\hat{X}A,\quad B\hat{D} =
\hat{B}D+\hat{X}B,  \\
C\hat{A} = \hat{C}A,\quad C\hat{B} = \hat{C}B,\quad D\hat{D} =
\hat{D}D,\quad C\hat{X} = p\hat{D}C,\quad D\hat{C} =
(1-p)\hat{D}C. 
\end{array}\label{eq:bulkalg2a} 
\end{equation}
There are some further relations related to the product form of
$\boldtau$ which forbids that products like
${\mathbf v}(\tau_{i-1},\tau_i) \otimes
{\mathbf v}(1-\tau_i,\tau_{i+1})$ occur in any physical quantity,
\begin{equation}
A\hat{D} = A\hat{C} = B\hat{A} = B\hat{B} =
B\hat{X} = C\hat{C} = C\hat{D} = D\hat{A} =
D\hat{B} = D\hat{X} =0. \label{eq:bulkalg2b}
\end{equation}
The boundary conditions from (\ref{eq:boundalg})
with the explicit values of the matrices $\L$ and $\R$ (\ref{eq:L&R}) become
\begin{equation}
\begin{array}{@{}rcl@{}}
\ln\hat{A} &=& (1-\alpha) \ln A \\
\ln\hat{B} &=& (1-\alpha) \ln B \\
\lo\hat{C} &=& \alpha \ln A + (1-p)\lo C \\
\lo\hat{D} &=& \alpha \ln B + \lo D \\
\ln\hat{X} &=& p \lo C
\end{array},\quad
\begin{array}{@{}rcl@{}}
A\rn &=& \hat{A} \rn + \beta \hat{B} \ro \\
B\ro &=& (1-\beta)\hat{B} \ro + \hat{X} \rn \\
C\rn &=& \hat{C} \rn + \beta \hat{D} \ro \\
D\ro &=& (1-\beta) \hat{D} \ro
\end{array}
\label{eq:boundalg2}
\end{equation}
Any solution to (\ref{eq:bulkalg2a}), (\ref{eq:bulkalg2b}) and
(\ref{eq:boundalg2}) thus automatically is a solution for the
stationary state via (\ref{eq:boundalg}) and  (\ref{eq:bulkalg}). It
is however not obvious that the cancellation mechanism of
(\ref{eq:boundalg}) and  (\ref{eq:bulkalg}) is appropriate for this
problem. Indeed, we will see that for the case of $p<1$ we will need a
slightly weakened version of (\ref{eq:bulkalg2a}) and (\ref{eq:boundalg2}).

By observation of explicit solutions for small system sizes we also have
inferred the following relations between the matrices,
\begin{equation}
\begin{array}{@{}rcl@{}}
DCB &=& \alpha \beta\left( (1-p) CB + DD + p\alpha\beta D \right),\\
BCB &=& \alpha \beta\left( AB + BD + p\alpha\beta B \right),\\
BCA &=& \alpha \beta\left( (1-p) BC + AA + p\alpha\beta A \right),\\
DCA &=& \alpha \beta (1-p) \left( DC + CA + p\alpha\beta C \right),\\
\end{array}
\label{eq:triple}
\end{equation}
with the following boundary conditions,
\begin{equation}
\begin{array}{@{}>{\dps}rc>{\dps}l>{\dps}rc>{\dps}l@{}}
\ln A &=& p\beta (1-\alpha) \ln & \quad B \ro &=& p\alpha \rn,\\
\ln B &=& p\beta \lo & \quad D \ro &=& p\alpha (1-\beta) \ro,\\
\lo CA &=& \alpha\beta (\ln A + (1-p) \lo C) & \quad DC \rn &=&
\alpha\beta (D \ro + (1-p) C \rn) ,\\
\lo CB &=& \alpha\beta (\ln B + \lo D) & \quad BC \rn &=& \alpha\beta
(A \rn + B \ro).\\
\end{array}
\label{eq:dbound}
\end{equation}
We believe (\ref{eq:triple}) and (\ref{eq:dbound}) to be true for
arbitrary system sizes. It turns out that these relations are
particularly convenient to obtain a representation. Using this
representation in (\ref{eq:bulkalg2a}) and (\ref{eq:boundalg2}) to
find the hatted matrices then gives an easy proof of the stationarity
of the Ansatz (\ref{eq:mpa}).

The relations (\ref{eq:bulkalg2b}) can be easily fulfilled by writing
\begin{equation}
\begin{array}{@{}>{\dps}rc>{\dps}l>{\dps}rc>{\dps}l@{}}
A &=& \A \otimes \eoo,\quad & B &=& \B \otimes \eot,\\[4mm]
C &=& \C \otimes \eto,\quad & D &=& \D \otimes \ett,
\end{array}
\end{equation}
and similarly for the hatted matrices
\begin{equation}
\begin{array}{@{}>{\dps}rc>{\dps}l>{\dps}rc>{\dps}l@{}}
\hat{A} &=& \hat{\A} \otimes \eoo,\quad & \hat{B} &=& \hat{\B} \otimes
\eot,\\[4mm] 
\hat{C} &=& \hat{\C} \otimes \eto,\quad & \hat{D} &=& \hat{\D} \otimes
\ett,\\[4mm]
\hat {X} &=& \hat{\X} \otimes \eoo.
\end{array}
\end{equation}
The boundary vectors are then written as
\begin{equation}
\begin{array}{@{}>{\dps}rc>{\dps}l}
\rn = \srn \otimes \left(\begin{array}{@{}c@{}} 1 \\ 0
\end{array}\right), & \quad \ro = \sro \otimes
\left(\begin{array}{@{}c@{}} 0 \\ 1 \end{array}\right) \\[6mm]
\ln = \sln \otimes (1,0), & \quad \lo = \slo \otimes (0,1).
\end{array}
\end{equation}

\section{Representations}
\label{se:rep}

\subsection{Representation for $1-p=(1-\alpha)(1-\beta)$}
Rajewski {\em et al.} \cite{Rajewsky:1998} have already remarked that
the product form (\ref{eq:mpa}) is exact for ordinary numbers instead
of matrices on the line $1-p=(1-\alpha)(1-\beta)$. Indeed, there exist a
one-dimensional representation given by
\begin{eqnarray}
\A=\beta (1-\alpha),\quad \B=1,\quad \C=\alpha\beta,\quad
D=\alpha(1-\beta) ,\nonumber\\
\hat{\A}=\beta(1-\alpha)^2,\quad \hat{\B}=1-\alpha,\quad
\hat{\C}=\alpha\beta(1-\alpha),\quad \hat{\D}=\alpha,\quad
\hat{\X}=\alpha p,
\end{eqnarray}
with
\begin{equation}
\sln = \beta,\quad \slo=1,\quad \srn = 1, \quad \sro = \alpha.
\end{equation}

\subsection{Representation for $p=1$}
In the case of deterministic dynamics in the bulk, $p=1$, a
two-dimensional representation of the subalgebra given by
(\ref{eq:bulkalg2a}) for general values of
$\alpha$ and $\beta$ can be found. The matrices $A,B,C$ and $D$ can be
easily read off from the solution in section \ref{se:matprod} and are
given by  
\begin{equation}
\begin{array}{@{}>{\dps}rc>{\dps}l>{\dps}rc>{\dps}l@{}}
\A &=& \left(\begin{array}{@{}cc} \beta(1-\alpha) & 0 \\ 0 & 0
\end{array}\right),\quad &
\B &=& \left(\begin{array}{@{}cc} \beta & 1 \\ 0 & \alpha
\end{array}\right),\\[4mm]
\C &=& \left(\begin{array}{@{}cc} \alpha\beta & -\alpha\beta \\ 0 &
\alpha\beta \end{array}\right),\quad &
\D &=& \left(\begin{array}{@{}cc} 0 & 0 \\ 0 & \alpha(1-\beta)
\end{array}\right).
\end{array} \label{eq:repp=1}
\end{equation}
The representation for the hatted matrices can then be found using the
algebra (\ref{eq:bulkalg2a}) and with its
boundary conditions (\ref{eq:boundalg2}) and is given by
\begin{equation}
\begin{array}{@{}>{\dps}rc>{\dps}l>{\dps}rc>{\dps}l@{}}
\hat{\A} &=& \left(\begin{array}{@{}cc} \beta(1-\alpha)^2 & 0 \\ 0 & 0
\end{array}\right),\quad &
\hat{\B} &=& \left(\begin{array}{@{}cc} \beta(1-\alpha) &
1-\alpha \\ 0 & 0 \end{array}\right),\\[4mm]
\hat{\C} &=& \left(\begin{array}{@{}cc} \alpha\beta(1-\alpha) & 0 \\ 0 & 0
\end{array}\right),\quad &
\hat{\D} &=& \left(\begin{array}{@{}cc} \alpha\beta & 0 \\ 0 & \alpha
\end{array}\right),\\[4mm]
\hat{\X} &=& \left(\begin{array}{@{}cc} \alpha\beta &
\alpha(1-\beta) \\ 0 & \alpha  \end{array}\right), 
\end{array} 
\end{equation}
with
\begin{equation}
\sln = (\beta,0),\quad \slo = (\beta,1), \quad \srn =
\left(\begin{array}{@{}c@{}} 1 \\ \alpha
\end{array}\right),\quad \sro = \left(\begin{array}{@{}c@{}} 0
\\ \alpha  \end{array}\right).
\end{equation}

\subsection{Infinite dimensional representation}
In appendix \ref{ap:infrep} it is explained that there exists a basis
$\{e_n,f_n\}$ on which the matrices take the following form,
\begin{equation}
\D=\B (1-p) = \alpha\beta q \left(
\begin{array}{@{}ccccc@{}}
q & 1 & 0 & 0 & \cdots \\
0 & q & 1 & 0 &  \\
0 & 0 & q & 1 &  \\
0 & 0 & 0 & q &  \\
\vdots & & & & \ddots  
\end{array}
\right),\quad
\A = \C = \alpha\beta q\left(
\begin{array}{@{}ccccr@{}}
q & 0 & 0 & 0 & \cdots\\
1 & q & 0 & 0 & \\
0 & 1 & q & 0 & \\
0 & 0 & 1 & q & \\
\vdots & & & & \ddots 
\end{array}
\right), \label{eq:infDmat}
\end{equation}
where $q=\sqrt{1-p}$. This representation does not lead to
divergent sums if 
\begin{equation}
\alpha,\beta > 1-\sqrt{1-p}.
\end{equation}
A representation that is valid for all values of $\alpha$ and $\beta$
can also be constructed, but this particular one will be useful for us
in the sequel. First of all we would like to diagonalize $E=A+B+C+D$ to
facilitate further calculations. To find the eigenvalues and
eigenvectors of $E$, we define the following vectors,
\begin{equation}
\zn{z} = \sum_{n=0}^\infty z^n e_n, 
\quad \zo{z} = \sum_{n=0}^\infty z^n f_n. \label{eq:zdef}
\end{equation}
It will also be convenient to define the following parameters,
\begin{equation}
a=\frac{p-\alpha}{\alpha q},\quad
b=\frac{p-\beta}{\beta q},
\end{equation}
so that the boundary vectors may be expressed as (see appendix
\ref{ap:infrep}), 
\begin{equation}
\rn = \kappa \frac{1-\beta}{1-p} \zn{b},\quad
\ro = \kappa \zo{b},
\end{equation}
where $\kappa$ is defined such that the normalization is given by
\begin{equation}
\ln V \rangle_0 = \beta,\quad \lo V \rangle_1 = \alpha.
\end{equation}
In appendix \ref{ap:eigen} it is shown that there exist vectors
$\ev{z}_\pm$ that are linear combinations of the vectors
defined in (\ref{eq:zdef}) which have the following properties, 
\begin{eqnarray}
E\ev{z}_+ &=& \Lambda_+(z) \ev{z}_+ =
\alpha\beta q \left( z+\frac{1}{z} + q
+ q^{-1}\right) \ev{z}_+,\\ 
E\ev{z}_- &=& \Lambda_- \ev{z}_- =
\alpha\beta q \left(q -  q^{-1}\right)
\ev{z}_-,
\end{eqnarray}
By writing the boundary vectors as linear combinations of
these eigenvectors, see (\ref{eq:leftboundary}) and
(\ref{eq:rightboundary}), the normalization
can be expressed as,
\begin{eqnarray}
Z_L &=& \left(\ln+\lo\right) E^{L-1} \left(\rn+\ro\right)
\nonumber\\ 
&=& -\tilde{\kappa} \oint_{|z|=1} \frac{\d z}{4\pi\i z}
(\Lambda_+(z)-\Lambda_-)\Lambda_+(z)^{L-1} K(z,a)K(z,b),
\label{eq:partsum1} 
\end{eqnarray}
where
\begin{equation}
K(z,c) = \frac{(z-z^{-1})}{(z-c)(z^{-1}-c)},\quad c=a,b,
\end{equation}
and
\begin{equation}
\tilde{\kappa} = \frac{1-p-(1-\alpha)(1-\beta)}{\alpha\beta(1-p)}
\;=\; \frac{1-a b}{p}. 
\end{equation}
Expression (\ref{eq:partsum1}) can be rewritten using the identities
in appendix \ref{ap:ident}. We then find 
\begin{equation}
Z_L = -\alpha\beta q
\frac{S_L(a)-S_L(b)}{p(\alpha-\beta)} = \frac{S_L(a)-S_L(b)}{a-b}, 
\end{equation}
where 
\begin{eqnarray}
S_L(c) &=& \frac{c}{p} \left(R_L(c) - \alpha\beta 
R_{L-1}(c) (q^2-1) \right) = \frac{c}{p}
\left(R_L(c)+p\alpha\beta R_{L-1}(c)  \right), \\ 
R_L(c) &=& \left(\alpha\beta q\right)^L \sum_{n=0}^L
\sum_{m=0}^{n} \left( q+q^{-1} -2 \right)^{L-n} \left(
\begin{array}{@{}c@{}} L \\ n \end{array} \right) \left(
\begin{array}{@{}c@{}} 2n-m \\ n \end{array} \right) \frac{m+1}{n+1}
(1+c)^m \label{eq:Rdef} \\
&\stackrel{(c<1)}{=}& -\oint_{|z|=1} \frac{\d z}{4\pi\i z}
\Lambda_+(z)^L K(z,c) (z-z^{-1}). \label{eq:intRL}
\end{eqnarray}
Note that the integral representation (\ref{eq:intRL}) for
$R_L(c)$ is only valid for $c<1$. Under this condition it can be
calculated to give (\ref{eq:Rdef}). Equation (\ref{eq:Rdef}) however
is valid for all values of $c$, which may be checked explicitly for
small system sizes or by using another infinite dimensional
representation for $c>1$.    

In deriving the phase diagram we will need the large $L$ behaviour of
$R_L(c)$. Expression (\ref{eq:Rdef}) for $R_L(c)$ is similar to that
of the ASEP with random sequential update, its asymptotics can be
calculated similarly \cite{Schuetz:1993,Derrida:1993}. By identifying
the terms that have the largest contribution to the sums we find that
$n \sim \sigma L$ and $m \sim 2/(1-c)$ for $c < 1$, $m \sim
\sqrt{2\sigma L}$ for $c=1$, $m \sim (c-1)\sigma L/c$ for $c >1$,
which gives, 
\begin{eqnarray}
R_L(c) &\approx& \frac{1}{\sqrt{\pi}} \left( \frac{2}{1-c} \right)^2
\Lambda_+(1)^L \frac{1}{(\sigma L)^{3/2}} 
\qquad{\rm for}\;c<1, \label{eq:Rc<1}\\ 
&\approx& \frac{2}{\sqrt{\pi}} \Lambda_+(1)^L \frac{1}{(\sigma
L)^{1/2}} 
\qquad{\rm for}\;c=1, \label{eq:Rc=1} \\
&\approx& (1-c^{-2}) \Lambda_+(c)^L 
\qquad{\rm for}\;c>1,\label{eq:Rc>1} 
\end{eqnarray}
where
\begin{equation}
\sigma = \frac{4}{2+q+q^{-1}},\quad \Lambda_+(1) = \alpha\beta
(1+\sqrt{1-p})^2,\quad \Lambda_+(a) = \alpha\beta
\frac{p^2(1-\alpha)}{\alpha(p-\alpha)}. 
\end{equation}
$\Lambda_+(b)$ is obtained from $\Lambda_+(a)$ by interchanging
$\alpha$ and $\beta$.

\section{Expressions for the current and density}

Using the algebra it is easy to derive the following expression for
the current $J_L = p\langle \tau_i (1-\tau_{i+1})\rangle_L$,
\begin{eqnarray}
J_L &=& p\frac{1}{Z_L} \left(\ln+\lo\right) E^iCE^{L-i-2}
 \left(\rn+\ro\right) \label{eq:Jdef}\\
&=& p\alpha\beta \left(\frac{Z_{L-1}}{Z_L} (1-2J_{L-1}) +
p\alpha\beta \frac{Z_{L-2}}{Z_L} (1-J_{L-2})\right) \label{eq:Jeq},
\end{eqnarray}
from which we find by induction,
\begin{equation}
J_L = p\alpha\beta \frac{Z_{L-1}}{Z_L}(1-J_{L-1}). \label{eq:Jeq2}
\end{equation}
The density profile is much harder to find from the algebra and our
strategy will be to express all correlation function in terms of the
eigenvectors of $E$. In doing so, the correlation functions are easily
expressed as integrals over the unit circle and can be calculated
exactly by the residue theorem or asymptotically via the saddle-point
method. We first demonstrate this for the current. Calculating the
action of $C$ on $\ev{z}_+$ using (\ref{eq:MatAct}) and
re-expressing it in the eigenvectors $\ev{z}_{\pm}$ we find,  
\begin{eqnarray}
C \ev{z}_+ &=& \alpha\beta q
\left[(1+zq)\zo{z} -
(1+z^{-1}q)\zo{z^{-1}} \right]\nonumber\\
&=& \alpha\beta \frac{\Lambda_+(z)}{\Lambda_+(z)-\Lambda_-}\left(
\ev{z}_+ - \ev{z}_-\right).
\end{eqnarray}
Inserting this into (\ref{eq:Jdef}) we find that
\begin{eqnarray}
J_L &=& -\frac{p\alpha\beta\tilde{\kappa}}{Z_L} \oint_{|z|=1}
\frac{\d z}{4\pi\i z} \Lambda_+(z)^{L-1} K(z,a)K(z,b) \nonumber\\
&=& \frac{\alpha \beta}{(a-b)Z_L} \left(
aR_{L-1}(a)-bR_{L-1}(b) \right), \label{eq:current}
\end{eqnarray}
which indeed fulfills (\ref{eq:Jeq2}).

In order to find the density profile we now calculate the two point
correlator $\langle \tau_i\tau_{i+1}\rangle_L$ which is given by
\begin{equation}
\langle \tau_i\tau_{i+1}\rangle_L = \frac{1}{Z_L} \left(\ln+\lo\right)
E^{i-1} D E^{L-i-1} \left(\rn+\ro\right).
\end{equation}
This can be given in an integral representation using
\begin{equation}
D\ev{z}_+ = \alpha\beta (1-p) \left[
z(1+zq)\zo{z} -
z^{-1}(1+z^{-1}q)\zo{z^{-1}} \right]. \label{eq:Doneig+}
\end{equation}
Re-expressing (\ref{eq:Doneig+}) as a linear combination of
eigenvectors using (\ref{eq:1identity}) it then follows that 
\begin{eqnarray}
\langle \tau_i\tau_{i+1}\rangle_L 
&=& -\frac{\alpha^2\beta^2 q
\tilde{\kappa}}{Z_L} \sum_{n=0}^\infty \oint _{|w|=1} \frac{\d
w}{2\pi\i w} \Lambda_+(w)^{i-1} K(w,a) (1+w^{-1}q) w^{-n}
\times \nonumber\\
&& \hphantom{-\frac{\alpha\beta q^{-1}
\tilde{\kappa}}{Z_L} \sum_{n=0}^\infty} \oint_{|z|=1} \frac{\d
z}{2\pi\i z} \Lambda_+(z)^{L-i-1} K(z,b) z (1+zq) z^n
\nonumber \\
&=& \frac{\alpha^2 \beta^2 \tilde{\kappa} q^2}{Z_L}
\sum_{m=0}^{L-i-1} R_{L-m-2}(a) R_{m}(b) + \frac{b q}{p} J_L
\label{eq:tpright} \\
&=& -\frac{\alpha^2 \beta^2 \tilde{\kappa} q^2}{Z_L}
\sum_{m=0}^{i-2} R_{m}(a) R_{L-m-2}(b) +
1-\frac{aq +q^2+1}{p}J_L. \label{eq:tpleft}
\end{eqnarray}
Here we have made use of the fact that the product
$\Lambda_+(w)^{i-1}\Lambda_+(z)^{L-i-1}$ can be written as
a sum in two ways. Equation (\ref{eq:tpright}) is useful for studying
the right boundary while (\ref{eq:tpleft}) is more suited for the left
boundary. The density profile is given by
\begin{equation}
\langle\tau_i \rangle_L = \langle\tau_i \tau_{i+1} \rangle_L + J_L/p 
\end{equation}
The easiest way to analyse the density profile is by looking at its
lattice derivative
\begin{equation}
t_L(i) = \langle\tau_{i+1}\rangle_L - \langle\tau_i \rangle_L =
-\frac{\alpha^2\beta^2\tilde{\kappa}q^2}{Z_L}
R_{i-1}(a)R_{L-i-1}(b). \label{eq:ddens}
\end{equation}
The value of the current and the asymptotic behaviour of $t_L(i)$ will
determine the phase diagram. 

\section{Phase diagram}
\label{se:phase}

\subsection{The case $p=1$}
\label{subse:p=1}
The two dimensional representation (\ref{eq:repp=1}) for this case
allow a simple evaluation of the current and density. The
current takes two different values corresponding to a low (LD) and a high
density (HD) region.

\begin{itemize}
\item{\bf Low density phase LD}

Here $\alpha < \beta$ and the current and density profile are given by
\begin{eqnarray}
J_- &=& \frac{\alpha}{1+\alpha}, \label{eq:p1LDcur}\\
\langle\tau_i\rangle_L &=& \frac{\alpha}{1+\alpha}
\left(1+\frac{1-\beta}{\beta} \e^{-r/\xi} \right),
\end{eqnarray}
where $\xi^{-1} = -\log (\alpha/\beta)$ and $r=L-i$. The density
profile is flat except near the right boundary where it falls of
exponentially from its maximum value $\langle\tau_L\rangle_L$ to the
bulk value.  

\item{\bf Transition line from LD to HD}

On this line $\alpha=\beta$. The current is still given by
(\ref{eq:p1LDcur}) but the density profile becomes linear,
\begin{equation}
\langle\tau_i\rangle_L =
\frac{\alpha}{1+\alpha} \left( 1+\frac{1-\alpha}{\alpha} \frac{i}{L}
\right).
\end{equation}

\item{\bf High density phase HD}
This phase is characterized by $\alpha > \beta$ and the current and
density can be obtained from those in the low density phase by the
particle hole symmetry (\ref{eq:phsym}). They are given by,
\begin{eqnarray}
J_+ &=& \frac{\beta}{1+\beta}, \label{eq:p1HDcur}\\
\langle\tau_i\rangle_L &=& \frac{1}{1+\beta}
\left(1-(1-\alpha) \e^{-i/\xi} \right),
\end{eqnarray}
where $\xi^{-1} = -\log (\beta/\alpha)$. Thus the density profile is
flat except near the left boundary where it increases exponentially from 
its minimum value $\langle \tau_1\rangle_L$ to its bulk value.
\end{itemize}

\subsection{General values of $p$}
The current (\ref{eq:current}) may take three different values
depending on the parameters $\alpha$ and $\beta$. These values
correspond to a low density, a high density and a so called maximum
current phase. The density profile in these phases will be
calculated and will give rise to a further discrimination of phases
within the low density and high density phase.

\begin{figure}[h]
\centerline{\begin{picture}(280,220)
\put(30,30){\epsffile{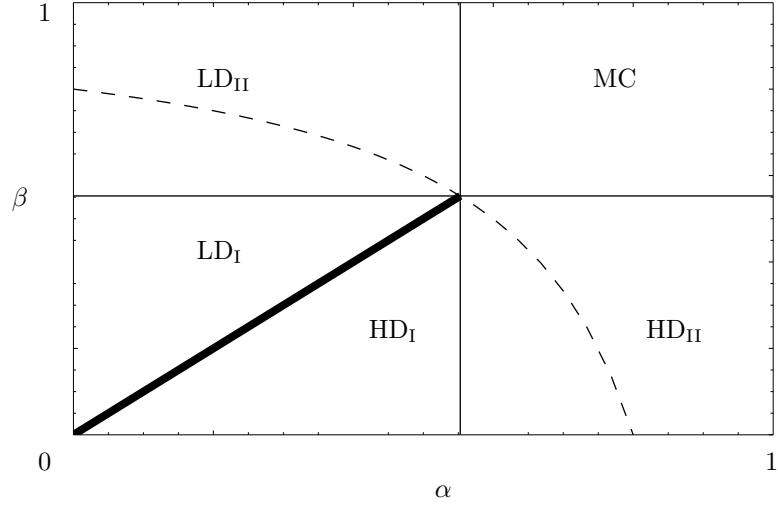}}
\put(230,165){MC}
\put(80,165){LD$_{\rm II}$}
\put(80,100){LD$_{\rm I}$}
\put(145,70){HD$_{\rm I}$}
\put(250,70){HD$_{\rm II}$}
\put(295,20){$1$}
\put(170,10){$\alpha$}
\put(20,20){$0$}
\put(10,120){$\beta$}
\put(20,190){$1$}
\end{picture}}
\caption{Phase diagram in the $\alpha-\beta$ plane. Phase boundaries
are at $\alpha,\beta=1-\sqrt{1-p}$ and at $\alpha=\beta$ where the
transition is discontinuous in the density. On the dashed
line, given by $1-p=(1-\alpha)(1-\beta)$, the mean field solution is
exact. \label{fig:phase}} 
\end{figure}

\begin{itemize}
\item{\bf Low density phase ${\rm LD_I}$}

This phase is chararcterized by the values $a>b>1$ or $\alpha<\beta<
1-\sqrt{1-p}$. The current and bulk density $\rho$ in this phase is
given by 
\begin{equation}
J_- = \frac{p\alpha\beta}{\Lambda_+(a)+p\alpha\beta} =
\frac{\alpha(p-\alpha)}{p-\alpha^2},\qquad \rho=1-J/\alpha.
\end{equation}
We find an exponential decay of the density profile with a length
scale
\begin{equation}
\xi^{-1} = \xi_a^{-1} - \xi_b^{-1},
\end{equation}
with
\begin{equation}
\xi_c^{-1} = -\log\left(\frac{\Lambda_+(1)}{\Lambda_+(c)}\right),\quad c=a,b.
\end{equation}
The slope of the density profile is given by (with $r=L-i$),
\begin{eqnarray}
t_L(i) &=& (b-b^{-1}) q J_- \left(1-\e^{-1/\xi}\right)
\e^{-(r-1)/\xi} \nonumber\\  
&=& \frac{(1-\alpha)(p-2\beta+\beta^2)}{(p-\alpha^2)
(1-\beta)} \left(1-\e^{-1/\xi}\right) \e^{-r/\xi}.
\end{eqnarray}
Since $b>1$ the slope of the density profile is positive and the
density approaches its bulk value from above.

\item{\bf Transition line from ${\rm LD_I}$ to  ${\rm LD_{II}}$}

On this line, where $b=1$ or $\beta=1-\sqrt{1-p}$, $\xi_b$ diverges
while $\xi_a$ remains finite. The bulk values are the same as in phase
${\rm LD_I}$, but the slope of the density profile now becomes 
\begin{eqnarray}
t_L(i) &=& \frac{2q J_-}{p\sqrt{\sigma\pi}} \left(
1-\e^{-1/\xi_a} \right) \e^{-(r-1)/\xi_a} r^{-1/2},\nonumber\\
&=& \frac{\alpha(p-\alpha)}{(p-\alpha^2)}
\frac{(1-p)^{1/4}}{\sqrt{\pi}(1-\sqrt{1-p})}  \left(
1-\e^{-1/\xi_a} \right) \e^{-(r-1)/\xi_a} r^{-1/2}.
\end{eqnarray}

\item{\bf Low density phase ${\rm LD_{II}}$}

This phase is determined by $a> 1 < b$ or $\alpha < 1-\sqrt{1-p} <
\beta$. Here the slope of the density profile still has a power law
correction to the exponential but the power is different. 
\begin{equation}
t_L(i) = \frac{q J_-}{p\sqrt{\pi \sigma}}
\frac{\Lambda_+(a) (\Lambda_+(a)-\Lambda_+(b))}
{(\Lambda_+(a)-\Lambda_+(1)) (\Lambda_+(b)-\Lambda_+(1))} \left(
1-\e^{-1/\xi_a} \right) \e^{-r/\xi_a} r^{-3/2} 
\end{equation}
The slope changes sign at the curve $a=b^{-1}$ or
$1-p=(1-\alpha)(1-\beta)$. This is the curve on which the mean field
solution is exact and a one dimensional representation of our algebra
exists. On this line the stationary state completely factorizes and
the density profile is flat.

\item{\bf Transition line from ${\rm LD_{II}}$ to the maximal current
phase ${\rm MC}$}

On this transition line the current and bulk value of the density are
as in the maximum current phase. The slope of the density profile on
this transition line is given by 
\begin{equation}
t_L(i) = -\frac{(1-p)^{1/4}}{4\sqrt{\pi}} (i/L)^{-1/2} r^{-3/2}.
\end{equation}
Near the right boundary the slope decays algebraically as
$r^{-3/2}$. Near the left boundary the slope of the density profile 
decays algebraically with a power of $1/2$ but the amplitude is of
order $1/L$. Thus up to order $1/L$ corrections the density profile
may be regarded as flat near the left boundary.

\item{\bf The maximal current phase ${\rm MC}$}

This part of the phase diagram is characterized by $a,b <1$ or
$\alpha,\beta > 1-\sqrt{1-p}$. In the maximal current phase the
current attains it maximum value which is first reached on its phase
boundaries. Its value and the bulk value of the density are given by 
\begin{equation}
J_{\rm max}=\frac{p\alpha\beta}{\Lambda_+(1)+p\alpha\beta} =
\frac{1}{2}(1-\sqrt{1-p}),\quad \rho=\frac{1}{2}.
\end{equation}
The slope of the density profile in this phase is given by
\begin{equation}
t_L(i) = -\frac{(1-p)^{1/4}}{4\sqrt{\pi}} i^{-3/2} (r/L)^{-3/2}.
\end{equation}
Since its slope is negative the density approaches its bulk value of
$\rho=1/2$ from above as $i^{-1/2}$ near the left boundary and from
below as $r^{-1/2}$ near the right boundary.

\item{\bf The high density phases ${\rm HD_I}$ and ${\rm HD_{II}}$}

The behaviour of the density profile in the high density phases and on
their phase boundaries can be obtained from those of the low density
phase by the particle hole symmetry, see equation (\ref{eq:phsym}),
\begin{equation}
\begin{array}{@{}>{\dps}rc>{\dps}l@{}}
\tau_{i-1} &\rightarrow & 1-\tau_r,\\
\alpha &\leftrightarrow & \beta.
\end{array} \label{eq:phsym2}
\end{equation}

\item{\bf Coexistence line}

This line is characterized by $a=b>1$ or
$\alpha=\beta<1-\sqrt{1-p}$. The length $\xi_a=\xi_b$ remains finite
but $\xi$ diverges. On this line one finds a linear profile with a
positive slope,
\begin{equation}
t_L(i) = \frac{p-2\alpha+\alpha^2}{(p-\alpha^2) L}.
\end{equation} 
\end{itemize}
In the limit of small rates, i.e. $\alpha=p\tilde{\alpha}$ and
$\beta=p\tilde{\beta}$ and $p\rightarrow 0$, we recover the results for
the ASEP with random sequential update
\cite{Schuetz:1993,Derrida:1993}. By taking $p\rightarrow 1$ the results
reduce to those derived in subsection \ref{subse:p=1}. Our
results are in perfect agreement with those of \cite{Evans:1998}.

In all phases and phase boundaries the current and bulk density $\rho$
satisfy the following relation which defines the fundamental diagram
\begin{equation}
J = \frac{1}{2}\left(1-\sqrt{1-4p\rho(1-\rho)}\right).
\end{equation}
Following Kolomeisky {\em et al.} \cite{Kolomeisky:1998}, we may
understand the phase diagram qualitatively by considering the domain
wall dynamics. In this picture two characteristic velocities are
important, the domain wall velocity and the collective velocity.
The collective velocity is the drift of the center of mass of a
momentary local perturbation of the stationary state and is related to
the current by,
\begin{equation}
V_{\rm coll} = \frac{\partial}{\partial \rho} J(\rho).
\end{equation}
This velocity changes sign at $\rho=1/2$ where the current takes its
maximum value. For positive domain wall velocity ($\beta > \alpha$)
and $\alpha < 1-\sqrt{1-p}$, an increase of the left boundary density
leads to an increase of the bulk density since $V_{\rm coll} >
0$. This happens until the left boundary density equals $1/2$, or
$\alpha = 1-\sqrt{1-p}$. At this point $V_{\rm coll}$ changes sign and
a perturbation will no longer spread into the bulk. The system is in the
maximal current phase and a further increase of the left boundary
density does not lead to an increase of the bulk density. For $\beta <
\alpha$ the system does not enter the maximal current phase
because of the negative domain wall velocity. The overfeeding however
still occurs, which implies that further increase of the left boundary
density beyond $1/2$ does not lead to changes of the characteristic
length scales in the high density phase. This is seen in the
divergence of the length scale $\xi_a$. 

The correlations for the ASEP with fully parallel dynamics are much
stronger than for other dynamics. This becomes apparent when
considering the relation between the length scales $\xi_{a,b}$ and the
curents in the high and low density phase. The lengths $\xi_{a,b}$ can be
written as,
\begin{equation}
\xi^{-1}_a = \xi^{-1}_{J_-},\quad \xi^{-1}_b = \xi^{-1}_{J_+},
\end{equation}
where
\begin{equation}
\xi^{-1}_J = -\log\left( \frac{J}{1-J} \frac{1-J_{\rm max}}{J_{\rm
max}} \right). \label{eq:Jlength}
\end{equation} 
This is in contrast to the random sequential and sublattice parallel
dynamics \cite{Schuetz:1993b,Hinrichsen:1996,Kolomeisky:1998} where
\begin{equation}
\xi^{-1}_J = -\log\left( \frac{J}{J_{\rm
max}} \right).
\end{equation}
In the latter cases this relation could be obtained directly by
considering the domain wall as a biased random walker. We have no
simple argument for the fluctations in the domain wall position that
leads to (\ref{eq:Jlength}) in the case of fully parallel dynamics.

\section{Conclusion}
We have presented a stationary state solution of an asymmetric simple
exclusion process with fully parallel dynamics. 
In the case of deterministic bulk dynamics the solution, 
obtained directly from the master equations, 
has the form of a product over two-dimensional matrices. 
In contrast to the ASEP's with other dynamics, the matrices
depend on two sites, instead of one. In the general case the
stationary state can still be written as a product over matrices, but of 
infinite size. The
stationarity of this product state can be proven by means of a
cancellation mechanism which is a bit weaker than in other cases.
We have calculated the exact phase diagram using an explicit 
representation of the matrix algebra. In this way we could via a
diagonalization procedure derive expression for the current and the
density profile with relative ease. 

The results are independent of, and agree with those of Evans {\em et 
al.}\cite{Evans:1998} which were obtained by means of a different 
Ansatz. They prove the strength and the flexibility of the matrix 
product Ansatz, though until recently the fully parallel dynamical models 
have resisted solution. Even when the resulting algebra is cubic (as in 
the present paper) or quartic (in \cite{Evans:1998}) a representation 
could be obtained. 
Of course, now that the formalism has been set 
up, many other properties of the stationary state can be calculated.
Instantaneous correlation functions are relatively straightforward.
As our representation includes probability distributions involving 
consecutive time steps, it is to be expected that the present formalism is 
capable in principle of producing time dependent correlation functions. 
A more difficult test of the formalism is the calculation of the
distribution of travelling times, for which it is necessary to follow a 
single particle as it flows through the system.

\section{Acknowledgment}
We thank M. Ernst and A. Wolters for useful discussions.
This work is part of the research program of the `Stichting voor 
Fundamenteel Onderzoek der Materie (FOM)' which is financially 
supported by the `Nederlandse organisatie voor Wetenschappelijk
Onderzoek (NWO)'.

\appendix

\section{Infinite dimensional representation}
\label{ap:infrep}

As an example for finding infinite dimensional matrices we explicitly
construct the representation used in the main text. First we choose
the following vectors as a basis, 
\begin{equation}
\{e_0,f_0,e_1,f_1,e_2,f_2,\ldots\},\label{eq:basis}
\end{equation} 
where
\begin{equation}
g_n = (\alpha\beta q)^{-n} (A-\alpha\beta (1-p))^n g_0,
\quad e_n = A g_n, \quad f_n = C g_n.
\end{equation}
Here $q=\sqrt{1-p}$ and we choose $g_0$ such that
\begin{equation}
Df_0 = \alpha\beta (1-p) f_0,\quad Bf_0 = \alpha\beta e_0.
\label{eq:act1}
\end{equation}
We then find the action of the matrices $A,B,C$ and $D$ on these
vectors from (\ref{eq:triple}) and (\ref{eq:dbound}). For example,
besides (\ref{eq:act1}) we find for $n\geq 1$,
\begin{eqnarray}
D f_n &=& \alpha\beta q ( f_{n-1} + q f_n),\\
B f_n &=& \alpha\beta  q^{-1} ( e_{n-1} + q e_n),
\end{eqnarray} 
so that $\D$ and $\B$ are indeed given by (\ref{eq:infDmat}). On the basis
(\ref{eq:basis}) the boundary vectors are given by
\begin{eqnarray}
\frac{1-p}{1-\beta} \langle {\mathcal V}_0| &=& \langle
{\mathcal V}_1| = \kappa \left(1, b,b^2,b^3,\ldots \right),\\  
\sln &=& \kappa \left(1,a, a^2,a^3, \ldots \right),\\
\slo &=& \frac{1}{p\beta} \sln \B = \kappa  q^{-1}
\left(\frac{\alpha q}{p}, 1,a,a^2,a^3, \ldots \right) , 
\end{eqnarray}
where
\begin{equation}
a=\frac{p-\alpha}{\alpha q},\quad
b=\frac{p-\beta}{\beta q},
\end{equation}
and
\begin{equation}
\kappa^2 = \frac{p(1-p-(1-\alpha)(1-\beta))}{\alpha(1-\beta)}.
\end{equation}
This representation does not lead to divergencies if $a,b <1$ or
$\alpha,\beta > 1-\sqrt{1-p}$. There are many possibilities in
choosing the set of basis vectors. We could for example define $g_0$
in a different way than was done in (\ref{eq:act1}). The
representation chosen here has some advantages which are exploited in
the main text. It is however possible to choose a representation which
is valid for all values of $a$ and $b$ (Evans {\em et al.} give an
explicit example of such a representation \cite{Evans:1998}). See
Derrida {\em et al.} \cite{Derrida:1993} for a similar discussion.

It turns out that for this representation we can find hatted
matrices satisfying the relations on the first line of
(\ref{eq:bulkalg2a}) but not those on the second line. It is however possible to relax the conditions of (\ref{eq:bulkalg2a}) a little in
the following way. Every matrix will be pre-multiplied by another
matrix. In particular $C$ or $D$ will be pre-multiplied by
$B$ or $D$ (or $\lo$ at the boundary). We 
therefore do not have to statisfy the relations on the second line of
(\ref{eq:bulkalg2a}) identically but only up to a term that vanishes
when acted on by $B$, $D$ or $\lo$. Since $\lo
-q\rangle\kern-.15em\rangle_1 = 0$ and $B\zo{-q} =  D\zo{-q} =0$, this
is the case if this 
term is a matrix of which the columns are multiples of
$\zo{-q}$. Thus, instead of the algebra obtained from
(\ref{eq:bulkalg}), we find a solution of the
algebra implied by 
\begin{eqnarray}
\lt\L\vecx = \lt\vecxt, && \vecx\otimes\R\vecxt\rt =
\vecx\otimes\vecx\rt, \label{eq:apboundalg}\\ 
\vecx\otimes\T\left[ \vecxt\otimes\vecx \right] &=&
\vecx\otimes\vecx\otimes\vecxt, \label{eq:apbulkalg} \\
\lt\T\left[ \vecxt\otimes\vecx \right] &=&
\lt\vecx\otimes\vecxt, \label{eq:apbulkboundalg}
\end{eqnarray}
A solution to this algebra still automatically gives rise to a
stationary state. We then find in addition to (\ref{eq:infDmat}),
\begin{eqnarray}
\hat{\A} &=& \hat{C}+p^2\alpha\beta \left(
\begin{array}{@{}ccccr@{}}
1 & (- q)^{-1} & (-q)^{-2} & (-q)^{-3} & \cdots\\
-q & 1 & (-q)^{-1} & (-q)^{-2} & \\
(-q)^2 & -q & 1 & (-q)^{-1} & \\
(-q)^3 & (-q)^2 & -q & 1 & \\
\vdots & & & & \ddots 
\end{array}
\right),\\
\hat{\C} &=& \alpha\beta\left(
\begin{array}{@{}ccccr@{}}
1-2p & -p^2 (-q)^{-1} & -p^2 (-q)^{-2} & -p^2(-q)^{-3} & \cdots\\
q & 1-2p & -p^2 (-q)^{-1} & -p^2(-q)^{-2} & \\
0 & q & 1-2p & -p^2 (-q)^{-1} & \\
0 & 0 & q & 1-2p & \\
\vdots & & & & \ddots 
\end{array}
\right),\\
\hat{\B} &=& \hat{\D} - p \alpha\beta \left(
\begin{array}{@{}ccccr@{}}
1 & 0 & 0 & 0 & \cdots\\
-q & 0 & 0 & 0 & \\
(-q)^{2} & 0 & 0 & 0 & \\
(-q)^{3} & 0 & 0 & 0 & \\
\vdots & & & & \ddots 
\end{array}
\right),\quad
\hat{\D}=\D+p\alpha\beta, \\
\hat{\X} &=& p\hat{\D} + p \alpha\beta \e^{2\lambda} \left(
\begin{array}{@{}ccccr@{}}
1 & 0 & 0 & 0 & \cdots\\
-q & 0 & 0 & 0 & \\
(-q)^{2} & 0 & 0 & 0 & \\
(-q)^{3} & 0 & 0 & 0 & \\
\vdots & & & & \ddots 
\end{array}
\right).
\end{eqnarray}

\section{Eigenvectors}
\label{ap:eigen}

It follows from a direct calculation that the actions of the matrices
on the vectors defined by (\ref{eq:zdef}) are given by
\begin{equation}
\renewcommand{\arraystretch}{1.4}
\begin{array}{@{}>{\dps}rc>{\dps}l}
A\zn{z} &=& \alpha\beta q\left( (z^{-1}+q)\zn{z}
-z^{-1}e_0\right),\\ 
B\zo{z} &=& \alpha\beta(1+z q^{-1}) \zn{z},\\
C\zn{z} &=& \alpha\beta q\left( (z^{-1}+q)\zo{z}
-z^{-1}f_0\right),\\ 
D\zo{z} &=& \alpha\beta q(q+z) \zo{z}.
\end{array} \label{eq:MatAct}
\end{equation} 
Taking linear combinations of different vectors $\zo{z}$ and $\zn{z}$ 
such that
terms with $e_0$ and $f_0$ drop out \cite{Karimipour:1998} we find the
following eigenvectors of $E$,
\begin{eqnarray}
\ev{z}_{\pm} &=& z\zn{z} -z^{-1}\zn{z^{-1}} + \eta_{\pm}(z)\zo{z} -
\eta_{\pm}(z^{-1}) \zo{z^{-1}},\\ 
\eta_+(z) &=& zq \frac{1+zq}{z+q},\quad
\eta_-(z) = -q.
\end{eqnarray}
The eigenvalues corresponding to these vectors follow easily from
(\ref{eq:MatAct}) and are,
\begin{equation}
\Lambda_+(z) = \alpha\beta q
\left(z+z^{-1}+q+ q^{-1}\right),\quad \Lambda_- =
\alpha\beta q \left(q -  q^{-1}\right).
\end{equation}From now on we take $|z|=1$. 
The following relations hold for $a<1$
and $b<1$ which are equivalent to $\alpha,\beta > 1-\sqrt{1-p}$.
\begin{eqnarray}
\left(\ln+\lo\right) \ev{z}_+ &=& \frac{\kappa}{p\beta}
\frac{(z-z^{-1})(\Lambda_+(z)-\Lambda_-)}{(z-a)(z^{-1}-a)},
\label{eq:leftboundary} \\
\left(\ln+\lo\right) \ev{z}_- &=& 0.
\end{eqnarray}
The vectors $\rn$ and $\ro$ can be expressed in the eigenvectors using
(\ref{eq:apident1}), from which we get,
\begin{equation}
-\oint_{|z|=1} \frac{\d z}{4\pi\i z} \frac{(z-z^{-1})}
{(z-b)(z^{-1}-b)} \ev{z}_+ = \kappa^{-1} \frac{1-p}{1-\beta}
\left(\rn+\ro-\beta^2 \kappa \zo{-q} \right) 
\label{eq:rightboundary}
\end{equation}
The third term on the right hand side is a null vector,
i.e. $\lo -q \rangle\kern-.15em\rangle_1=0$ and
$E\zo{-q}=0$, and does not enter the calculations. 

\section{Identities}
\label{ap:ident}

The following identity which is frequently used throughout this paper
can be conveniently calculated (or looked up in
\cite{Gradshteyn:1980}) by writing the denominator of the
integrand as a sum of two geometric series and using the residue
theorem,
\begin{equation}
c^{k-1} = -\oint_{|z|=1} \frac{\d z}{4\pi\i z} \frac{(z^k-z^{-k})
(z-z^{-1})} {(z-c)(z^{-1}-c)},\quad c<1. \label{eq:apident1}
\end{equation}
In a similar fashion the following integral can be calculated for $c<1$,
\begin{eqnarray}
\lefteqn{-\oint_{|z|=1} \frac{\d z}{2\pi\i z} (2+z+z^{-1})^L z^n
\frac{z-z^{-1}}{(z-c)(z^{-1}-c)}} \nonumber\\ 
&=& \sum_{k=0}^{L-n} \renewcommand{\arraystretch}{1} \sum_{k=0}^{L+n}
\left[ \binom{2L}{k} c^{L+n-k-1} - \binom{2L}{k} c^{L-n-k-1} \right],
\label{eq:moments}
\end{eqnarray}
for $0<n\leq L$. Specialising to $n=1$ and rewriting the terms in the
sum we find that
\begin{equation}
-\oint_{|z|=1} \frac{\d z}{4\pi \i z} \frac{(z-z^{-1})^2
(2+z+z^{-1})^N} {(z-c) (z^{-1}-c)}\stackrel{(c <1)}{=} \sum_{p=0}^N
\left( \begin{array}{@{}c@{}} 2N-p \\ N \end{array}
\right)\frac{p+1}{N+1} (1+c)^p. 
\end{equation}
Another important identity that we use to express vectors in terms of
the eigenvectors of $E$ is
\begin{eqnarray}
&&\oint_{|z|=1} \frac{\d z}{4\pi\i z}
\frac{\alpha\beta  q^{-1}}{\Lambda_+(z)-\Lambda_-}
\left(\ev{z}_+ -\ev{z}_- \right)
\left((1+z^{-1}q) \lsub{1}{\langle\kern-.15em\langle z|} -
(1+zq) \lsub{1}{\langle\kern-.15em\langle z^{-1}|} \right)
\nonumber\\  
&=& \oint_{|z|=1} \frac{\d z}{4\pi\i z} \left(\frac{z}{z+q}
\zo{z} - \frac{1}{1+zq} \zo{z^{-1}} \right)
\left((1+z^{-1}q) \lsub{1}{\langle\kern-.15em\langle z|} -
(1+zq) \lsub{1}{\langle\kern-.15em\langle z^{-1}|} \right)
\nonumber\\ 
&=& I_1 - (1-q^2) \zo{-q}
\kern-.1em\langle\kern-.15em\langle - q^{-1}|. \label{eq:1identity}
\end{eqnarray}
This again can be simply evaluated using the residue theorem and the
fact that $q < 1$.

\end{document}